\documentclass[doublecol]{epl2} 

\usepackage{graphicx}
\usepackage{color}
\usepackage{amssymb,amsfonts,amsmath}
\usepackage[english]{babel}
\title{Modulation of current through a nanopore induced by a charged 
globule: implications for DNA-docking}
\shorttitle{Current through a nanopore induced by a charged globule}

\author{
Mauro Chinappi\inst{1} \and 
Carlo Massimo Casciola\inst{2,3} \and
Fabio Cecconi\inst{4} \and 
Umberto Marini Bettolo Marconi\inst{5} \and
Simone Melchionna\inst{6}}

\shortauthor{M. Chinappi \etal}

\institute{
\inst{1} Center for Life Nano Science@Sapienza, Istituto Italiano di
Tecnologia, Via Regina Elena 291, 00161, Roma, Italia

\inst{2} Dipartimento di Ingegneria Meccanica e Aerospaziale,
Sapienza Universit\`a di Roma, Via Eudossiana 18, 00184, Roma, Italia

\inst{3} Center for Life Nano Science@Sapienza, Istituto Italiano di
Tecnologia, Via Regina Elena 291, 00161, Roma, Italia

\inst{4} CNR-Istituto dei Sistemi Complessi, Via Dei Taurini 19, 
00185 Roma, Italy 

\inst{5} Scuola di Scienze e Tecnologie, Universit\`a di Camerino, 
Via Madonna delle Carceri, 62032 Camerino, INFN, sez. di Perugia, Italy 

\inst{6} CNR-IPCF, Consiglio Nazionale delle Ricerche, Dipartimento di Fisica, Universit\`a Sapienza, Piazzale A. Moro 2, 00185 Roma, Italy 
}
\pacs{66.90.+r}{Other topics in nonelectronic transport properties of condensed matter}
\pacs{66.10.Ed}{Ionic conduction}
\pacs{61.25.H-}{Macromolecular and polymers solutions; polymer melts}

\abstract{
The passage of DNA through a nanopore can be effectively decomposed into two 
distinct phases, docking and  actual translocation. In experiments each 
phase is characterized by a distinct current signature which allows 
the discrimination of the two events. However, at low voltages 
a clear distinction of the two phases is lost.
By using numerical simulations we clarify how the 
current signature associated to the docking events depends on the applied voltage.  
The simulations show that at small voltage the DNA globule enhances 
the pore conductance due to an enrichment of charge carriers. 
At high voltage, the globule drains substantial charge carriers from 
the pore region, thereby reducing the overall conductance. The results
provide  a new interpretation to the experimental data on conductance
and show how docking interferes with the translocation 
signal, of potential interest for sequencing applications.
}

\begin{document}

\maketitle

\section{Introduction}
In the last decade, nanopore-based biosensing has become a burgeoning research field
thanks to the impressive burst in the capability to 
fabricate devices based on solid state 
\cite{russo2012atom,storm2003fabrication}
and biological pores 
\cite{hall2010hybrid,mohammad2012engineering,laszlo2013detection}.
The working principle of the device is, in essence, simple. 
The nanopore connects two chambers containing an electrolyte solution.
Under an electric voltage, ions 
%and any charged macromolecule 
migrate from one chamber 
to the other, with a conductance that depends on the details of the pore and the 
electrolytic solution. 
Since a translocating macromolecule alters the ionic flux, its
passage can be detected and possibly the local nature of its monomers
can be read off.
Translocation of DNA is the most studied process
\cite{kasianowicz1996characterization}
%with the goal of realizing a personalized genome sequencing device.
being a promising technology for low cost/high throughput DNA analysis, 
its realization is still confronted with challenges
%~\cite{storm2005fast,kowalczyk2012measurement,fyta2006multiscale,schneider2012dna}.
~\cite{pennisi2014dna,kowalczyk2012measurement,fyta2006multiscale,schneider2012dna}.
In parallel, a separate set of studies have focused on protein and 
polypeptide translocation,
with the goal of structural characterization
\cite{rodriguez2013multistep,bacci2012role,bacci2013protein,
madampage2012nanopore,cressiot2012protein}. 

Over the pore extension, the DNA segment has a virtually rod-like conformation and
during the threading process the ionic current increases for low salt ($< 0.4\,M$ KCl)
and decreases for high salt concentration~\cite{chang2004dna,fan2005dna}
as compared to the open pore (so called free-pore in the following) case.
The accepted explanation for the current reduction 
is that at high salinity the ionic motion is sterically hampered by the presence of DNA
that reduces the effective pore section~\cite{smeets2006salt}.
At low salinity instead, 
the counterions cloud is more diffused and the charge carriers more mobile, an effect that
prevails over the steric hinderance, increasing the conductance 
as compared to the free-pore condition\cite{chang2004dna,fan2005dna}.

Recently, Kowalczyk and Dekker \cite{kowalczyk2012measurement}
and Vlassarev and Golovchenko\cite{vlassarev2012trapping}
reported of a new type of hybrid event with two distinct signals, a
current decrease followed by a current increase.
Interestingly, the amount of current reduction is comparable to that of current enhancement 
and the transition between the two regimes takes place on a timescale shorter than 
the translocation event.
The experiments were performed in low salt conditions and, hence, 
the increase in current was associated to the threading of DNA, while
the decrease of current was interpreted as due to the presence of the DNA at the pore entrance.
In fact, since the electric field is highly focused
in the pore and access regions, the modulation of ionic current takes
place only when DNA is in close proximity to the pore, in the so-called docked configuration.
In addition, current reduction was attributed to the frictional forces 
exerted by the docked DNA on the solution~\cite{kowalczyk2012measurement}.
However, the physical mechanism responsible for decreasing the ionic current 
remains elusive. In fact, while frictional forces can reduce electrokinetic transport,
the transition between docked and threading DNA should be gradual.
If friction has a leading role, a decrease of conductance
should persist during DNA threading, since large portions of the DNA coil
sit on both the cis and trans chambers of the device.
%%%%%%%%%  Fig1 %%%%%%%%%%%%%%%%%%%%%%%%%%%%%%%%%%%%%%%%%%%%%%%
%\begin{figure*}
 \begin{figure}[]
	\centering
	\includegraphics[angle=-0,width=0.48\textwidth]{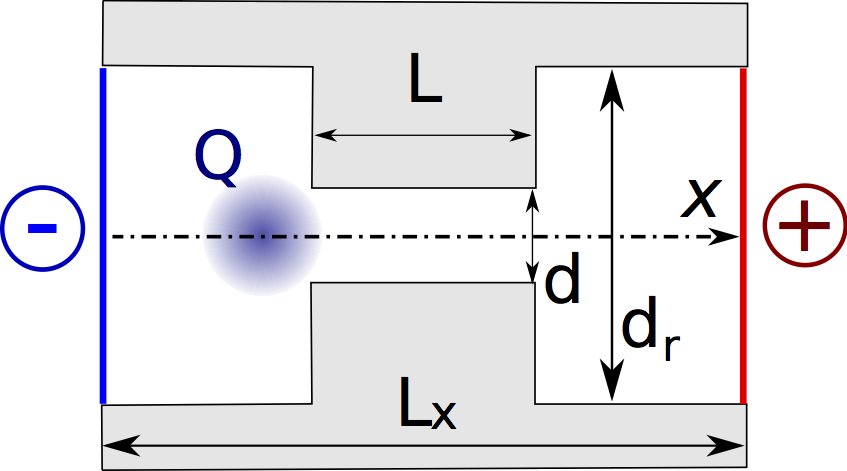}
	\caption{A) Sketch of the system. A cylindrical 
                 pore of length $L$ and diameter $d$ connects two reservoirs 
                 of diameter $d_r$, with total length $L_x = L + 2 L_r$, and
                 aligned along the $x$ axis.
                 Bulk concentrations and the electric potential are imposed at the
                 left $(-)$ and right $(+)$ boundaries.
                 An immobile globule of total
                 negative charge $Q$ is located near the pore entrance.
                 Note that the sketch does not reflect the actual  
                    aspect ratio of the system, see Fig. \ref{fig:currex}
		}
	\label{fig:geom}
 \end{figure}
%\end{figure*}
%%%%%%%%%  End Fig1 %%%%%%%%%%%%%%%%%%%%%%%%%%%%%%%%%%%%%%%%%%%
%
%%%%%%%%%  Fig1 %%%%%%%%%%%%%%%%%%%%%%%%%%%%%%%%%%%%%%%%%%%%%%%
\begin{figure}
	\centering
	\includegraphics[width=0.5\textwidth]{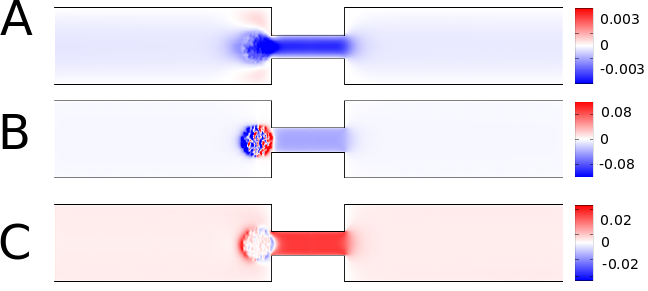}
	\caption{The current fields $J^n$, $J^+$ and $J^-$, for the neutral, positive and negative 
species, on the y-z plane. 
A: $J_x^n$, driven by electro-osmosis, on average directed 
($− \leftarrow +$), 
B: $J_x^+$, on average directed ($− \leftarrow +$) 
and C: $J_x^-$, on average directed ($− \rightarrow +$). 
Currents are reported in units of $\rho^\alpha_{bulk}v_T$ \label{fig:fields}}
	\label{fig:currex}
\end{figure}
%%%%%%%%%  End Fig1 %%%%%%%%%%%%%%%%%%%%%%%%%%%%%%%%%%%%%%%%%%%

%
Assessing the contribution to conductance stemming from a docked DNA is relevant
to correctly discriminate the ionic signal arising from the translocating nucleotides.
Due to the comparable intensity of ionic currents associated to docked and translocating
conformations, subtracting the systematic contribution 
originating from partially translocated DNA globules is particularly useful 
in the low voltage regime,
when translocation is slow and therefore the base pair content is amenable to detection.
The present study aims at understanding the origin of current
modulation by employing computer simulations in a model representation of the system.  
We consider an idealized version of the device 
with DNA modelled as a charged globule docked near the pore entrance.

Our results reproduce qualitatively the modulation of current at varying voltages 
as observed in experiments.
In addition, they suggest an alternative cause for the observed modulation
that is based entirely on electrokinetic effects. In particular,
i) a substantial depletion of ions takes place within the pore due to the
asymmetric charge drainage exerted by the globule,
% on the solution,
ii) the current decrease can be explained in terms of charge rearrangements,
without the need for considering excluded volume or hydrodynamics interactions
between DNA and the pore,
iii) at low voltage the current increases as compared to the free-pore case, 
a regime that has not been detected by experiments.
We further suggest that at low voltages the current increase observed in experiments
should be assigned to hybrid docked/translocating configurations.\\

%%%%%%%%%%%%%%%%%%%%%%%%%%%%%%%%%%%%%%%%%%%%%%%%%%%%%%%%%%%%%%%
\section{System set-up and simulation method}

We consider a cylindrical nanopore of length $L$ and diameter $d = 0.3\,L$ 
connecting two reservoirs of length $L_r = 3\,L$ and diameter 
$d=L$, (Fig. \ref{fig:geom}).
The origin of the coordinate system is located at the center of the pore
and the x-axis coincides with the pore axis. 
A potential difference $\Delta V$ is applied between a
positive and a negative electrode, placed at $x = \pm 3.5\,L$, respectively.
The fluid is composed of three
species: positive charge carriers, negative charge carriers and a neutral 
species~\cite{marini2012charge,marconi2013effective}.
The density fields are indicated
as $\rho^+$, $\rho^-$ and $\rho^{n}$ respectively. 
The electrolytic solution is assumed to be neutral and homogeneous 
far away from the pore, with $\rho^+_{bulk}=\rho^-_{bulk}$  
at the two open boundaries ($x = \pm3.5\,L$).
The Bjerrum and Debye lengths are set to $l_B = 0.035\, L$  
and $\lambda_D = 0.05\, L$, respectively. 
These values are chosen in order to 
map the conditions employed 
in the experiments~\cite{kowalczyk2012measurement}
\footnote{In Ref.\cite{kowalczyk2012measurement} 
$L \simeq 20\, nm$ and $d \simeq 6\, nm$ 
while, $l_B = 0.7\, nm$ and $l_D = 10\, nm$,  corresponding to a $0.15\,M$ 
solution of a monovalent salt in water.}.

The role of the wall charge in the translocation experiment is two-fold: 
to enhance the electro-osmotic current by altering the amount of free charge carriers 
via the Donnan effect  and to introduce additional interactions with the macromolecule 
\cite{cressiot2012protein, pedone2009translo}.
In the present work we neglect the presence of surface wall charge in order 
to keep the number of simulation parameters to a bare minimum and
to isolate the effect of the docking globule on the ionic current. 
Such approximation allows capturing the leading effects in the modulation 
of the ionic conduction, while the presence of additional free charge carriers 
can be accounted for by modifying the bulk molarity. 
Secondly, as we are interested in docking conditions, 
the surface charge is expected to play a minor role on the DNA globule and charge 
redistribution outside the pore region.

The presence of DNA docked at the pore entrance is mimicked by
a spherical distribution of negative charges is positioned in a fixed location 
near the pore entrance, the total globule charge being $Q$ (Fig. \ref{fig:geom}).
% Details on the mathematical model and the Lattice Boltzmann formulation 
% are reported in the appendix. \\

%%%%%%%%%  Fig3 %%%%%%%%%%%%%%%%%%%%%%%%%%%%%%%%%%%%%%%%%%%%%%%
\begin{figure}
	\centering
	\includegraphics[width=0.48\textwidth]{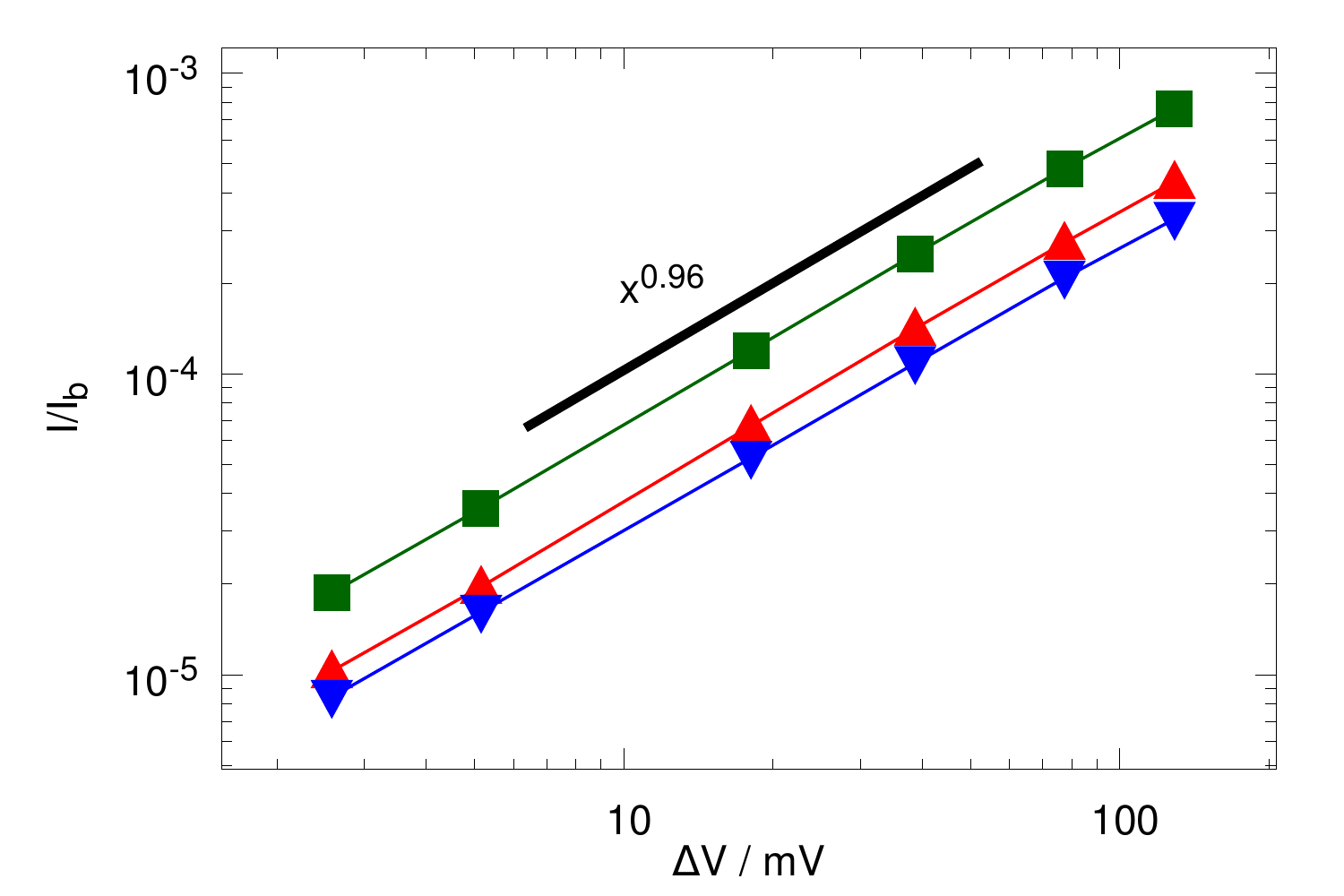}
	\caption{Positive ($\vert I^+ \vert$, triangles) and 
                 negative 
                 ($\vert I^- \vert$, inverted triangles) current intensities as 
                 functions of the applied voltage.
                 Squares refer to the electrical 
                 current intensity ($I_c = \vert I^+ - I^- \vert$). 
                 A non linear dependence of the electrical current 
                 intensity is apparent 
                 ($I_c \sim \Delta V^\alpha$ with $\alpha = 0.93 \pm 0.01$).
                 The current is represented in units of 
                 $I_b  = \pi (d_r/2)^2 \rho^+_{bulk} / v_T$.
	\label{fig:current}}
\end{figure}
%%%%%%%%%  End Fig3 %%%%%%%%%%%%%%%%%%%%%%%%%%%%%%%%%%%%%%%%%%%

%%%%%%%%%  Fig3 %%%%%%%%%%%%%%%%%%%%%%%%%%%%%%%%%%%%%%%%%%%%%%%
\begin{figure}[]
	\centering
	\includegraphics[width=0.44\textwidth]{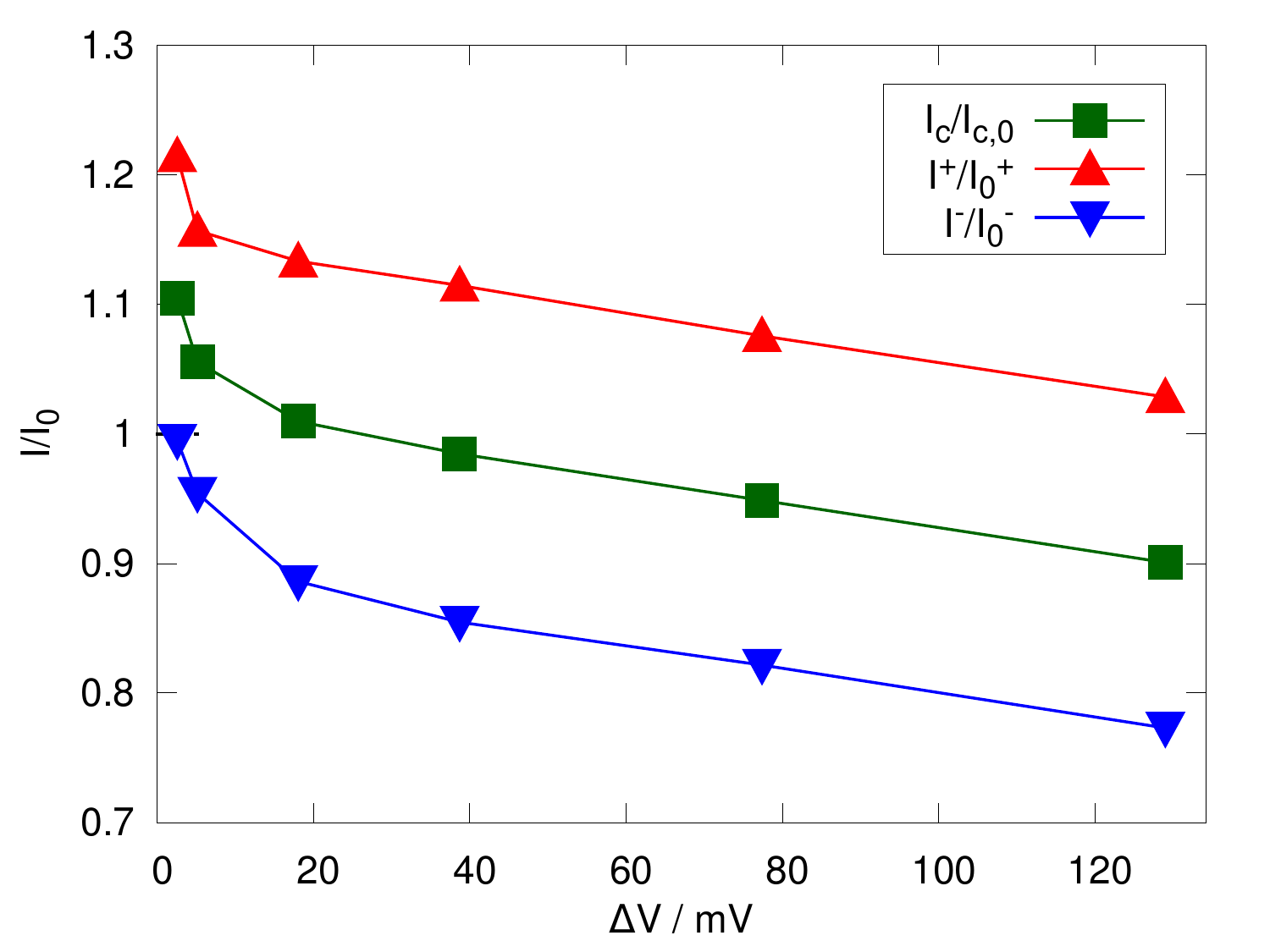}
	\caption{
        Current in the presence of the globule
        at the pore entrance normalized by the value 
        in absence of the globule, as a function of the applied voltage.
        Blue inverted triangle and red triangles refer to negative and positive         carriers 
        respectively, while green squares refer to the total electric current. 
	\label{fig:IsuI0}
}
\end{figure}
%%%%%%%%%  End Fig3 %%%%%%%%%%%%%%%%%%%%%%%%%%%%%%%%%%%%%%%%%%%

The motion of the electrolytic solution is described by the density,
velocity and electric fields of each species,
having number densities $\rho^\alpha=\rho^n,\rho^+,\rho^-$, equal unit mass,
valence $z^\alpha= 0,1, −1$,
and equal mobilities.
We adopt the continuum approach, implicitly assuming that the motion of individual molecules 
is not relevant for the problem at hand.
Several studies indicate that for simple fluids this is a good 
description down to the nanoscale~\cite{chinappi2008mass,bocquet2010nanofluidics}.
The electrokinetic flow is simulated by means of the Lattice Boltzmann Method,
which reproduces the advection-diffusion dynamics for three species,
a neutral one representing the solvent, and two charged species for the electrolytes,
in the presence of electrostatic interactions
\cite{marini2012charge,marconi2013effective,melchionna2011electro}. 
Simulations were performed with the parallel multiphysics software 
MUPHY\cite{bernaschi2009muphy}.

Atomistic simulations and recent experiments indicate that 
the slip length $L_s$ for water on smooth surfaces hardly exceeds
$1\, nm$~\cite{chinappi2010intrinsic,attard2013more}, therefore
no-slip boundary conditions are applied for each fluid species 
at the solid wall ~\cite{chinappi2010intrinsic,attard2013more}.
The materials usually employed for fabricating the nanopore are 
hydrophilic, thus allowing to safely  disregard slip effects 
at the scale of interest.
For electrostatics, the Poisson equation,
$\nabla^2 \phi=-{1\over \epsilon }\left(e \rho^+ - e \rho^- 
+ \theta\right)$, 
% + Q \theta({\bf x} - {\bf R})\right)$, 
where 
$\theta$ is the globule 
charge distribution,
is solved with appropriate boundary conditions.  
The tangential component of  the electric field 
is continuous at the liquid-solid interface.
Given the large disparity in the dielectric constants for water ($\epsilon \simeq 80$)
and the wall ($\epsilon \simeq 3.9$  for silicon oxide and $\simeq 7.9$ for silicon nitride),
we impose Neumann boundary conditions,
$\nabla \Phi\cdot \hat{\bf n} = 0$.
where 
$\hat{\bf n}$ is the unit vector orthogonal to the wall.
Molecular Dynamics simulations have shown that the dielectric constant of water
confined in a cavity of diameter $5\,nm$ is $71$, close to the 
value in the bulk \cite{senapati2001dielectric}. 
Hence, given the size of our system ($d = 6\,nm$), 
a minor alteration of the dielectric constant is expected
and we employed $\epsilon = 80$ in the whole system.

The docked globule is represented as a system of regularized point-wise charges,
each carrying a unit charge $0.1\,e$, and
distributed uniformly within a sphere of diameter $d$, i.e. the same diameter
of the pore. 
The interaction strength is given by the Bjerrum length, $l_B = e^2/4\pi \epsilon k_BT$,
which is taken to be $0.7\, nm$, corresponding to liquid water conditions at ambient temperature.
In order to resolve the double layer structure and
avoid compressibility artifacts arising from the numerical method, 
we set the spacing of the computational grid to $0.02\, nm$. 
Fully developed flow is assumed at the inlet and at the outlet sections,  
each having a prescribed electrical potential 
($\phi_{in} = 0$ and $\phi_{out} = \Delta V$). Dirichlet boundary conditions 
enforce the bulk densities of each species at the inlet and outlet,
in particular, $\rho^+_{in} = \rho^+_{out} = \rho^-_{in} = \rho^-_{out}$.

For the Lattice Boltzmann method, the electrolytic solution is represented by three distribution 
functions, $f^\alpha(x,v,t)$, representing the probability of having a particle of 
species $\alpha$ at position $x$, with velocity $v$ and at time $t$. 
The species evolve according to the equation
\begin{equation}
\partial_t f^\alpha + \nabla \cdot v f^\alpha = \omega(f^\alpha_{eq} - f^\alpha)
+{e z^\alpha\over m} \nabla \phi \cdot {\partial\over \partial v} f^\alpha
\equiv C^\alpha[\{f\}]
\end{equation}
where $f^\alpha_{eq} = \left[{1\over 2\pi v_T^2}\right]^{3/2} \rho^\alpha 
e^{-{(v-u)^2\over 2 v_T^2}}$ is the Maxwell-Boltzmann local equilibrium,
$v_T$ the thermal velocity, 
$u={\sum_\alpha \rho^\alpha u^\alpha \over \sum_\beta \rho^\beta}$ is the barycentric
fluid velocity and $u^\alpha$ the species velocity. The quantity $\omega$
controls the rate of relaxation towards local equilibrium and is related to the 
kinematic viscosity $\nu$ as $\nu=v_T^2(1/\omega-1/2)$.  
The distributions $f^\alpha$ are discretized 
in space over a Cartesian mesh and expanded in velocity space over a Hermite basis set. 
By Hermite projection, the distributions are replaced by 
$f^\alpha(x,v,t)\rightarrow f^\alpha_p$ and analogously 
$C^\alpha(x,v,t)\rightarrow C^\alpha_p$.
The dynamics is generated according to the updating scheme
\begin{equation}
f_p^\alpha(x+c_p,t+1)-f_p^\alpha(x,t)=C_p^\alpha(x,t)
\end{equation}
where $c_p$ represents a set of discrete speed that connects a mesh point to its 
neighbors. The species density is computed as
$\rho^\alpha = \sum_p f^\alpha_p$
and the species current as
$J^\alpha \equiv \rho^\alpha u^\alpha = \sum_p c_p f^\alpha_p$.
Example of the obtained current fields are shown in Fig.~\ref{fig:currex}.

%%%%%%%%%%%%%%%%%%%%%%%%%%%%%%%%%%%%%%%%%%%%%%%%%%%%%%%%%%%%%%%%

\section{Results} 

The current intensity of the species $\alpha$ is defined
as $I^\alpha = \int_{A(x)} \rho^\alpha u_x^\alpha dA  $ where $A(x)$ is the local section 
perpendicular to the pore axis, $\rho^\alpha$ 
is the number density of a given species and $u_x^\alpha$ the x-component 
of the velocity field. The total ionic current is $I_c = I^+ -I^-$.   
As a preliminary check, we verified that in stationary flow conditions
the system is  globally neutral, and that 
in the absence of the globule, 
 the number of positive and negative carriers is exactly the same.
Given the geometrical symmetry and the 
identical valence and mobility of the charged species
the positive charge flow (electrodes $(+)\rightarrow (-)$) 
exactly balances the flow of negative carriers
(electrodes $(-)\rightarrow (+)$), so that no electro-osmotic flow, i.e. 
the net motion of the neutral specie, is present. 
Moreover, the system response is found to be purely {\sl Ohmic} and the 
ionic currents are individually proportional to the applied voltage $V$ 
(data not shown).
In the following, we refer to quantities
in the absence of the globule (free-pore case) with the subscript $0$.
 
The presence of the globule results in an abundance of 
positive carriers necessary to maintain the condition of global electroneutrality.
The unbalance between positive and negative carriers 
reflects in a difference between $\vert I^+ \vert$ and $\vert I^- \vert$,
see Fig.\ref{fig:current}.

To appreciate the role of the applied voltage, 
it is instructive to compare the data in the presence and in the absence 
of the globule. 
Fig. \ref{fig:IsuI0} reports the ratio between the current 
intensities for total globule charges $Q=100\,e$ and $Q=0$. 
In the explored voltage range the presence of the globule 
at the pore entrance leads to an increase
of positive carrier current, $I^+/I^+_0 > 1$ (red triangles in Fig. \ref{fig:IsuI0})
 and to a 
depletion of the negative carrier one, 
$I^-/I^-_0 < 1$ (blue inverted triangles in Fig. \ref{fig:IsuI0}).
This modulation is caused by the 
already mentioned abundance of positive carriers.

\begin{figure}[]
	\centering
	\includegraphics[width=0.44\textwidth]{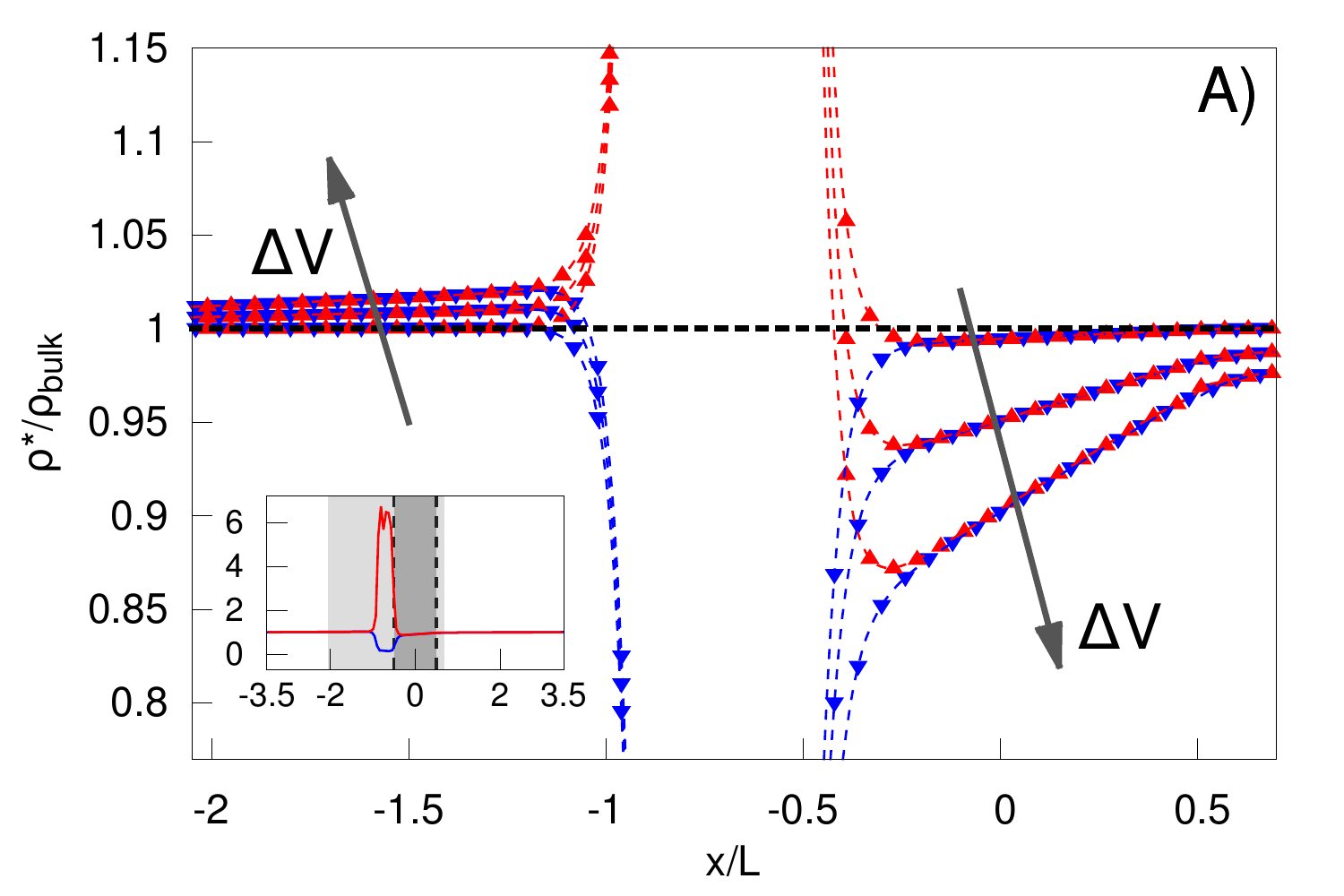}
	\includegraphics[width=0.44\textwidth]{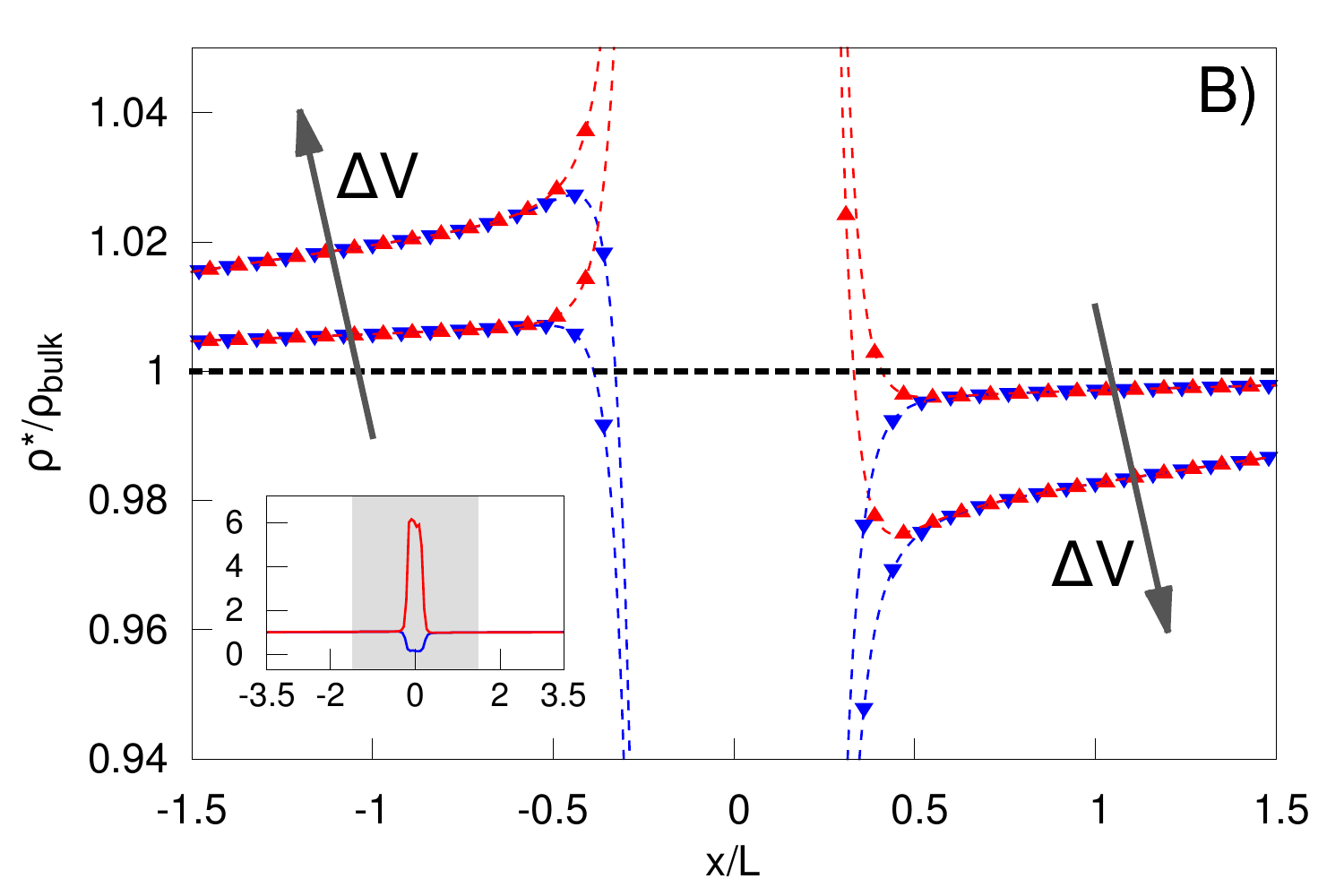}
	\caption{A: density profiles of positive (red triangles) and 
                 negative  (blue inverted triangles) species
                 normalized by the bulk value, for applied  
                 voltage $\Delta V = 5.2,\,38.8,\,77.5 \,mV$ 
                 (the arrows indicate the trend at increasing voltage). 
                 Inset: zoom-out of the profiles,
                 with the light shaded area showing the data reported in the main graph
                 and the dark shaded area the pore region.
                 B: same as in the upper panel but in absence of                 the pore ($\Delta V = 18.1,\, 77.5 \,mV$). 
                 }
	\label{fig:dens}
\end{figure}

A second observation is that both $I^+/I^+_0$ and $I^-/I^-_0$ decrease 
as the applied voltage $\Delta V$ increases,
consistently with the depletion of carriers in the pore region
as the voltage increases, see Fig.~\ref{fig:dens}A.
Interestingly at low voltages $I_c/I_{c,0} >0$, i.e. the presence
of the charged globule results in an increase of the current, while
at high voltage $I_c/I_{c,0} < 0$.
The conductance modulation as a function of $\Delta V$ is 
interpreted by considering
the simple case of a globule placed in the middle of 
a cylindrical tube of length $L_x = L + 2 L_r$ and diameter 
$d_r = L$, i.e. 
a single cylinder with the same diameter of the reservoir with no 
interposed nanopore, spanning the whole length of the system. 
At zero voltage, the charge distribution is symmetric 
around the globule and, outside the Debye layer
(i.e. at distance $>3\lambda_D$ from the globule), 
both $\rho^+$ and $\rho^-$ recover the bulk value imposed at the two open boundaries. 
Under a finite voltage, the counterion cloud surrounding the globule is asymmetrically distorted 
towards the negative electrode. The counterion density is therefore larger towards the negative electrode (charge surplus)
than towards the positive electrode (charge depletion),  
see Fig.~\ref{fig:dens}B.  
Clearly, outside the Debye layer the solution recovers local electro-neutrality 
($\rho^+ = \rho^-$),
with an antisymmetric arrangement of charges with respect to the globule position induced by 
the applied voltage.

The presence of the pore breaks the fore-aft symmetry and results in 
a substantial carrier depletion inside the pore (i.e. towards the positive 
electrode with respect to the globule). Such depletion is one order of magnitude more intense 
than the corresponding increase towards the negative electrode, 
see Fig.~\ref{fig:dens}A. The pore induces an imperfect
electrostatic screening as compared on the opposite side of the globule.
The effect increases with $\Delta V$ and hence, at high voltages, 
the carrier density inside the pore is strongly depleted by the charged globule.

In essence, the docked globule produces
two concomitant effects:
i) the global increase in the available charge carriers 
and ii) the local depletion of carriers inside the pore.
At low $\Delta V$, the depleted pore plays a marginal role on 
the overall conduction and the leading effect is the increase 
of positive carriers, hence $I_c > I_{c,0}$. 
At high $\Delta V$, the depleted pore dominates
on the increase of the available counterions, hence $I_c < I_{c,0}$. 

%%%%%%%%%  Fig4 %%%%%%%%%%%%%%%%%%%%%%%%%%%%%%%%%%%%%%%%%%%%%%%
%%%%%%%%%  End Fig4 %%%%%%%%%%%%%%%%%%%%%%%%%%%%%%%%%%%%%%%%%%%

\begin{figure}[]
	\centering
	\includegraphics[width=0.44\textwidth]{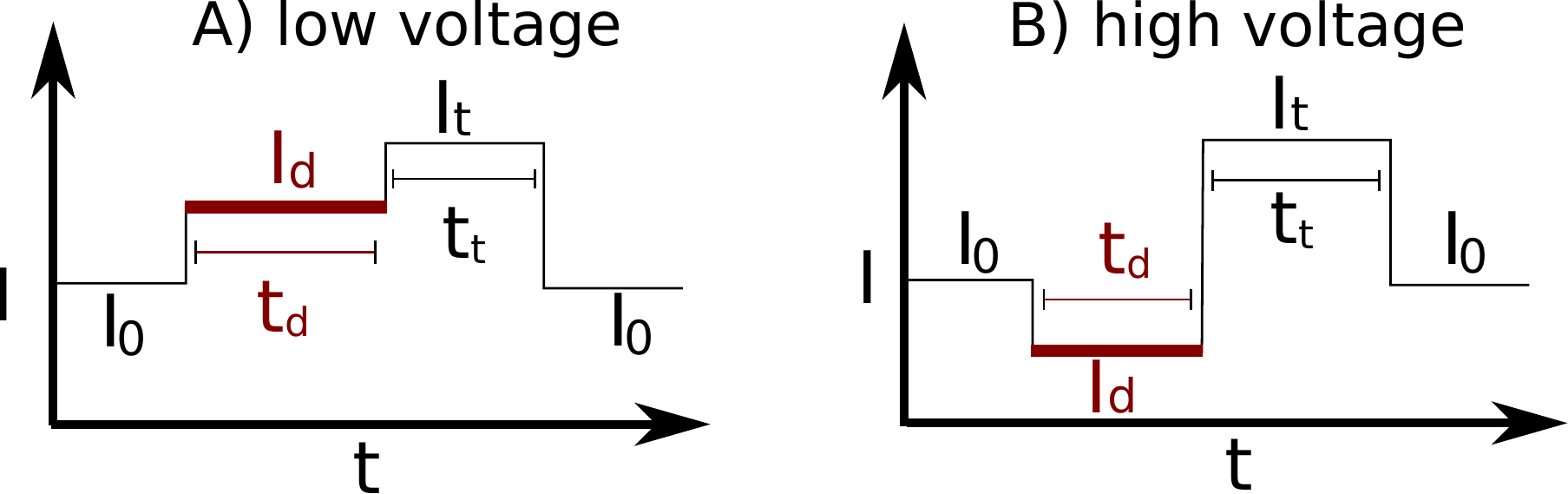}
	\includegraphics[width=0.44\textwidth]{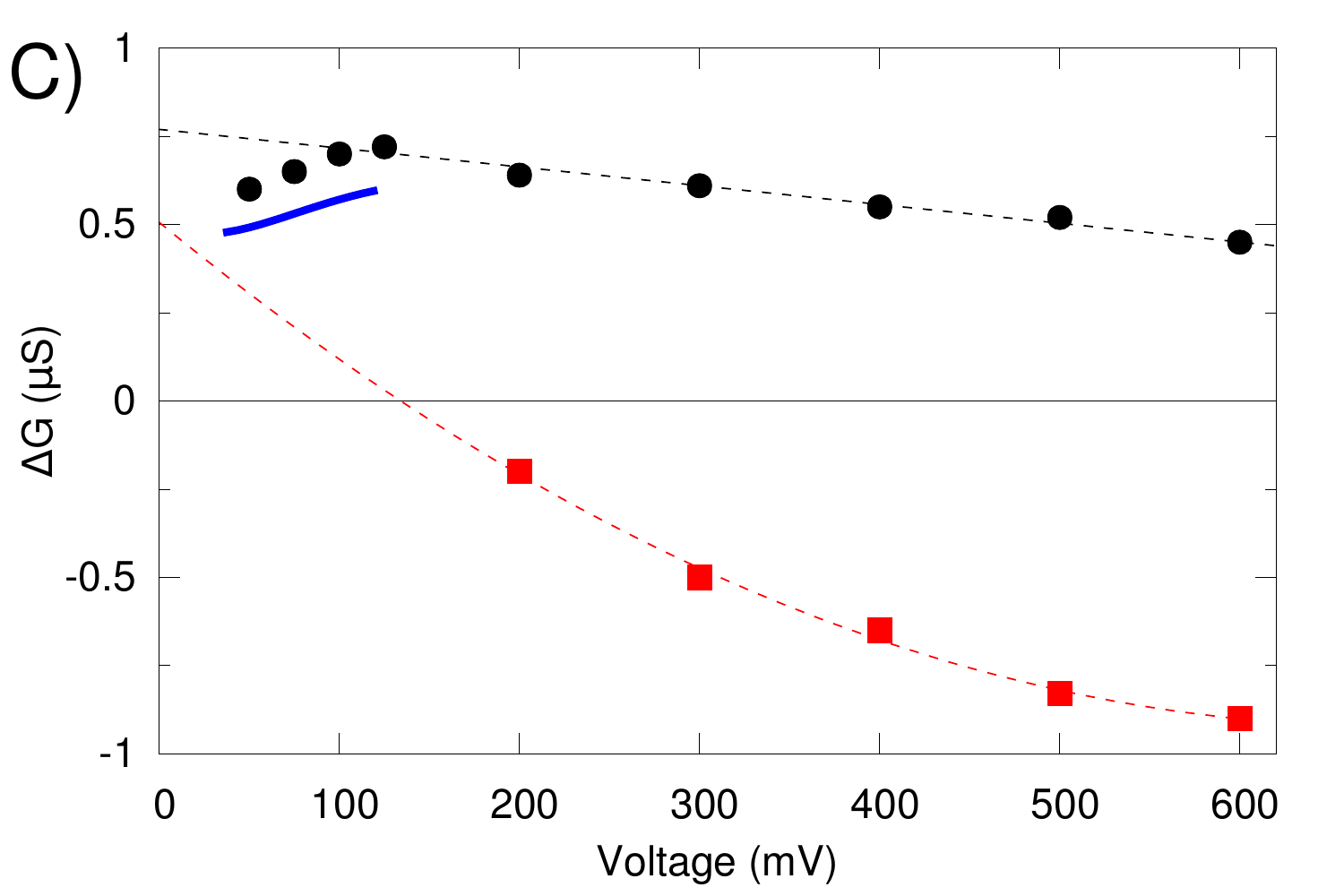}
	\caption{Upper panels: low (a) and high (b) voltage scenarios. 
                 At low voltage both docking and translocation events increase the current. 
                 Lower panel: experimental differential conductance
                 \cite{kowalczyk2012measurement},
                 associated to docking (red squares)
                 and translocation (black circles).
                 The dashed curves are fits of the experimental data 
                 for $200\,mV <\Delta V <600 \,mV$.  The blue line 
                 is the new estimate, eq.~(\ref{eq:dg}),
                 for $50\,mV <\Delta V <125 \,mV$.
}
	\label{fig:last}
\end{figure}

It is illustrative to relate the results of the present
simulations with experimental findings of 
Kowalczyk and Dekker~\cite{kowalczyk2012measurement}.
We remind that these authors reported the intensity and the 
average duration of the current modulations, as shown in Fig.~\ref{fig:last}C.
The current decrease was associated to docking while the current increases to translocation events.  
Let us now focus on the current decrease
and the differential conductance, defined as $\Delta G \equiv (I_c - I_{c,0}) / \Delta V$.
The experimental docking conductance decreases monotonically in the range 
$200 \,mV < \Delta V < 600 \,mV$
(red squares in Fig. \ref{fig:last}C)
in qualitative agreement with our numerical result in the
voltage range where $I_c/I_{c,0} < 1$. 
At small voltages ($\Delta V < 200 mV$), the experiments do not detect any current decrease, 
whereas the simulations show that $I_c/I_{c,0} > 1$. 
Moreover, experiments show an optimal enhancement at $V\simeq 125 \,mV$ 
(black circles  in Fig.~\ref{fig:last}C).
The apparent discrepancy between experiments and simulations at low voltages can
be reconciled according to the following argument.
For $200 \,mV <\Delta V < 600 \,mV$ the experiments
detect hybrid events, with a clear-cut separation between reduction/docking
(of duration $t_d$) and enhancement/translocation (of duration $t_t$) events.
Two distinct signals, called $I_d$ and $I_t$ respectively, are measured 
as averages over the time intervals $t_d$ and $t_t$ (Fig. \ref{fig:last}B). 
At low voltage, the simulations indicate that the ionic current is enhanced as compared
to the free-pore case. Since both the docking and the translocation events 
feature an increase in conductance (Fig. \ref{fig:last}A) 
the docking events are hardly distinguishable  from the translocation ones.
In particular, as voltage diminishes, the docking events become slower than the translocating 
ones, as reported in~\cite{kowalczyk2012measurement}.

To estimate the conductance for low $\Delta V$, we average the experimental 
data of~\cite{kowalczyk2012measurement}
over docking and translocating events, as
\begin{equation}
\Delta G 
=  
\frac
{\Delta G_{d} t_d + \Delta G_t t_t} 
{t_d + t_t} \ .
\label{eq:dg}
\end{equation}
where $\Delta G_{d}$, $\Delta G_t$, $t_d$ and $t_t$ 
depend individually on the applied voltage.
We evaluate the conductance in the range $ 50 \,mV <\Delta V < 125 \,mV$ 
by using for $t_d$ and $t_t$ the expressions of ref. \cite{kowalczyk2012measurement},
and fit the experimental
data in the interval $200\,mV  < \Delta V < 600 \,mV$ to
extrapolate the differential conductances $\Delta G_d$ 
and $\Delta G_{t}$ at low voltage.
In passing, we note that the differential conductance due to docking events is zero for
$\Delta V\simeq 130 \,mV$. This value is comparable to the numerical one ($\simeq 30\,mV$),
notwithstanding the approximations involved, in particular 
regarding the large disparity of the globule size
between simulation and experiments and the absence of pore wall charge.
The final result of the averaging of eq. (1) is reported in 
Fig. \ref{fig:last}C (blue solid line), 
which exhibits a good agreement with the experimental result, thereby lending good confidence
in the proposed interpretation.

In conclusion, the present work underscores the importance of electrokinetic effects on 
DNA translocation experiments. In particular, it indicates that DNA docked 
near the pore entrance can either enhance or reduce the pore conductance 
at different applied voltages as compared to the free-pore case.
Such non trivial phenomenon depends on the imperfect electrostatic screening 
that depletes the pore region from charge carriers. 
On the other hand, as the macromolecule begins threading, the conductance
is dominated by the DNA thread in the pore region that
increases and compensates for the charge drainage exerted by the 
DNA coil at the pore entrance. 
In the future, it will be interesting to investigate 
the role played by the charged wall and how the interplay between 
coil induced drainage and thread induced enrichment 
affects ionic conductance during DNA translocation.

%%%%%%%%%  Fig5 %%%%%%%%%%%%%%%%%%%%%%%%%%%%%%%%%%%%%%%%%%%%%%%
%%%%%%%%%  End Fig5 %%%%%%%%%%%%%%%%%%%%%%%%%%%%%%%%%%%%%%%%%%%

\acknowledgments

Computing time at the CINECA supercomputing center (ISCRA grants FLEXPROT and 
NAPS) is kindly acknowledged.

%%%%%%%%%%%%%%%%%%%%%%%%%%%%%%%%%%%%%%%%%%%%%%%%%%%%%%%%%%%%%%%%
\bibliography{traslo}

%\end{thebibliography}

\end{document}